\documentclass[preprint,showpacs,preprintnumbers,amsmath,amssymb,floatfix]{revtex4-1}

\usepackage{graphicx}
\usepackage{dcolumn}
\usepackage{bm}

\begin{document}

\title{Fission dynamics at low excitation energy}

\author{Y.~Aritomo and S.~Chiba}

\affiliation{Research Laboratory for Nuclear Reactors, Tokyo Institute of Technology,
Ookayama, Meguro-ku, Tokyo, 152-8850, Japan}%




\begin{abstract}

The origin of mass asymmetry in the fission of uranium at a low excitation energy is
clarified by a trajectory analysis of the Langevin equation.
The positions of the peaks in the mass distribution of fission fragments
are mainly determined by fission saddle points
originating from the shell correction energy.
The widths of the peaks, on the other hand, result from a shape
fluctuation around the scission point
caused by the random force in the Langevin equation.
We found that a random vibration in the oblate direction of fissioning fragments
is essential for the fission process.
According to this picture, fission does not occur with continuous stretching in the prolate direction,
similarly to that observed in starch syrup.
This is expected to lead to a new viewpoint of fission dynamics and the splitting mechanism.

\end{abstract}




\maketitle



\section{Introduction}

Since the  discovery of the nuclear fission of uranium in 1938 \cite{harn39,meit39},
the principle of this phenomenon has been studied owing to its scientific interest,
and its application to the supply of power was realized soon after its discovery.
However, asymmetric fission remained a puzzle when nuclei were described using the analogy of
a liquid drop \cite{bohr39}.
The origin of the asymmetry in the mass distribution of fission fragments (MDFFs) is still unclear,
 although it appears to be connected to the nuclear structure of the fission fragments.
Many theoretical models have been applied to nuclear fission in an attempt to explain its
mechanism.

A dynamical approach using the Langevin equation can be used to investigate the
time evolution of the nuclear shape during fission.
In our previous study \cite{arit13}, this approach was applied to the fission
of U and Pu at low excitation energies while taking into account the shell structure of the nuclei.
In the calculation, we obtained an asymmetric MDFF that agreed with
experimental data as well as the total kinetic energy (TKE) of the fission fragments.
Our approach simultaneously gave accurate descriptions of the measured MDFF and the TKE.

In the present work, on the basis of our previous study \cite{arit13}, we attempt to clarify the origin
of the asymmetric MDFF of U at a low excitation energy.  With this aim, Langevin trajectories were
analyzed precisely and the time evolution of nuclear shapes was explored.
We found the factors determining the positions and widths of the peaks in the MDFF:
the former is mainly related to the positions of the fission saddle points, which are caused
by the shell correction
energy, and the latter is related to the thermal fluctuation caused by the random force in the Langevin equation
around the scission point.

In addition, we obtained a new interpretation of the mechanism of fission dynamics.
The fluctuation of a fissioning nucleus in the oblate direction causes a compact configuration at the scission point.
By comparing the TKE of the fission fragments obtained experimentally and by calculation,
we confirmed that such a configuration is realized there.
This leads to the picture that fission does not occur in the manner of starch syrup, which grows with a very small neck radius with continuous stretching.

In this paper, we present these new interpretations of fission dynamics at a low excitation energy.
The paper is organized as follows.
In Sec.~II, we describe the framework of the model in detail.
In Sec.~III, we reveal the origin of mass-asymmetric fission for $^{236}$U at
$E^{*}=20$ MeV by analyzing the trajectories, and we discuss the potential landscape.
In Sec.~IV, we investigate the fission dynamics and the configuration at the scission point.
The role of the friction tensor in fission dynamics is described in Sec.~V.
In Sec.~VI, the TKE of the fission fragments is discussed.
We present a summary of this study and further discussion in Sec.~VII.

\section{Model}

We use the fluctuation-dissipation model and
employ Langevin equations \cite{arit04} to investigate the dynamics of the fission process.
The nuclear shape is defined by the two-center
parametrization \cite{maru72,sato78}, which has three deformation parameters, $z_0, \delta$, and $\alpha$ to serve as collective coordinates:
$z_{0}$ is the distance between two potential centers,
while
$\alpha=(A_{1}-A_{2})/(A_{1}+A_{2})$
is the mass asymmetry of the
two fragments, where
$A_{1}$ and $A_{2}$ denote the mass numbers of heavy and light fragments \cite{arit04}. The symbol
$\delta$ denotes the deformation of the fragments, and
is defined as
$\delta=3(R_\parallel-R_\perp)/(2R_\parallel+R_\perp)$, where $R_\parallel$ and $R_\perp$
are the half length of the axes
of an ellipse in the $z_0$ and $\rho$ directions of the cylindrical coordinate,
respectively, as shown in Fig.~1 in Ref. \cite{maru72}.
We assume in this work that each fragment has the same $\delta$.  This constraint
should be relaxed in the future work since the deformations of
the heavy and light fragments in the fission of U region are known to be different from each other.
The deformation parameters $\delta$ and $\beta_{2}$ are related to each other as
\begin{equation}
\beta_{2}=\frac{\delta}{\sqrt{\frac{5}{16\pi}}(3-\delta)}.
\end{equation}
Notice that $\delta < 1.5$ since  $R_\parallel > 0$ and $R_\perp > 0 $.
In order to reduce the computational time, we employ the coordinate $z$ defined as
$z=z_{0}/(R_{CN}B)$, where $R_{CN}$ denotes the radius of a spherical compound nucleus
and $B$ is defined as $B=(3+\delta)/(3-2\delta)$.
We use the neck parameter $\epsilon=0.35$, which is recommended in Ref. \cite{sato78}
for the fission process.
The three collective coordinates may be abbreviated as $q$, $q=\{z,\delta,\alpha\}$.

For a given value of a temperature of a system, $T$,
the potential energy is defined as a sum of the liquid-drop (LD) part,
a rotational energy and a microscopic (SH) part;
\begin{equation}
V(q,\ell,T)=V_{\rm LD}(q)+\frac{\hbar^{2}\ell(\ell+1)}{2I(q)}+V_{\rm SH}(q,T),
 \label{vt1}
\end{equation}
\begin{equation}
V_{\rm LD}(q)=E_{\rm S}(q)+E_{\rm C}(q).
\end{equation}
\begin{equation}
V_{\rm SH}(q,T)=E_{\rm shell}^{0}(q)\Phi (T),
\label{XevKK}
\end{equation}
\begin{equation}
\Phi (T)=\exp \left(-\frac{aT^{2}}{E_{\rm d}} \right).
\label{XevKK2}
\end{equation}
Here, $V_{\rm LD}$ is the potential energy calculated with the finite-range liquid drop model,
given as a sum of of the surface energy $E_{\rm S}$ \cite{krap79} and the Coulomb energy $E_{\rm C}$.
$V_{\rm SH}$ is the shell correction energy evaluated by Strutinski method from the single-particle
levels of the two-center shell model.  The shell correction have a temperature dependence expressed
by a factor $\Phi (T)$,
in which $E_{\rm d}$ is the shell damping energy chosen to be 20 MeV \cite{igna75} and
$a$ is the level density parameter.
At zero temperature ($T=0$), the shell correction energy reduces to that of the two-center shell
model $E_{\rm shell}^{0}$.
The second term on the right hand side
of Eq. (\ref{vt1}) is the rotational energy
for an angular momentum $\ell$ \cite{arit04},
with a moment of inertia at $q$, $I(q)$.

The multidimensional Langevin equations \cite{arit04} are given as
\begin{eqnarray}
&&\frac{dq_{i}}{dt}=\left(m^{-1}\right)_{ij}p_{j},\nonumber \\
&&\frac{dp_{i}}{dt}=-\frac{\partial V}{\partial q_{i}}
                 -\frac{1}{2}\frac{\partial}{\partial q_{i}}
                   \left(m^{-1}\right)_{jk}p_{j}p_{k}
-\gamma_{ij}\left(m^{-1}\right)_{jk}p_{k}
                  +g_{ij}R_{j}(t),
\label{lange1}
\end{eqnarray}
where $i = \{z, \delta, \alpha\}$ and
$p_{i} = m_{ij} dq_{j}/dt$
is a momentum conjugate to coordinate $q_i$.
The summation is performed over repeated indices.
In the Langevin equation,
$m_{ij}$ and $\gamma_{ij}$ are the shape-dependent collective inertia and the
friction tensors, respectively.
The wall-and-window one-body dissipation
\cite{bloc78,nix84,feld87}is adopted for the friction tensor which can describe the
pre-scission neutron multiplicities and total kinetic energy of fragments simultaneously\cite{wada93}.
A hydrodynamical inertia tensor is adopted with the Werner-Wheeler approximation
for the velocity field \cite{davi76}.
The normalized random force $R_{i}(t)$ is assumed to be that of white noise, {\it i.e.},
$\langle R_{i}(t) \rangle$=0 and $\langle R_{i}(t_{1})R_{j}(t_{2})
\rangle = 2 \delta_{ij}\delta(t_{1}-t_{2})$.
The strength of the random force $g_{ij}$ is given by Einstein relation $\gamma_{ij}T=\sum_{k}
g_{ik}g_{jk}$.

The temperature $T$ is related with the intrinsic energy
of the composite system as $E_{\rm int}=aT^{2}$, where
$E_{\rm int}$ is calculated at each step of a trajectory calculation as
\begin{equation}
E_{\rm int}=E^{*}-\frac{1}{2}\left(m^{-1}\right)_{ij}p_{i}p_{j}-V(q,\ell,T=0).
\end{equation}
%


The fission events are determined in our model calculation by identifying the different
trajectories in the deformation space.
Fission from a compound nucleus is defined as the case that
a trajectory overcomes the scission point on the potential energy surface.

%
%

\section{Origin of mass asymmetric fission} 

In our previous study \cite{arit13}, we investigated the fission of $^{236}$U at the excitation
energy $E^{*}=20$ MeV, and we obtained the MDFF, which agreed with the experimental data indicating
that mass-asymmetric fission is dominant.
In the present paper, by analyzing the trajectories in our model, we attempt to clarify the origin
of the mass-asymmetric fission events of uranium at low excitation energies.

\begin{figure}
\centerline{
\includegraphics[height=.80\textheight]{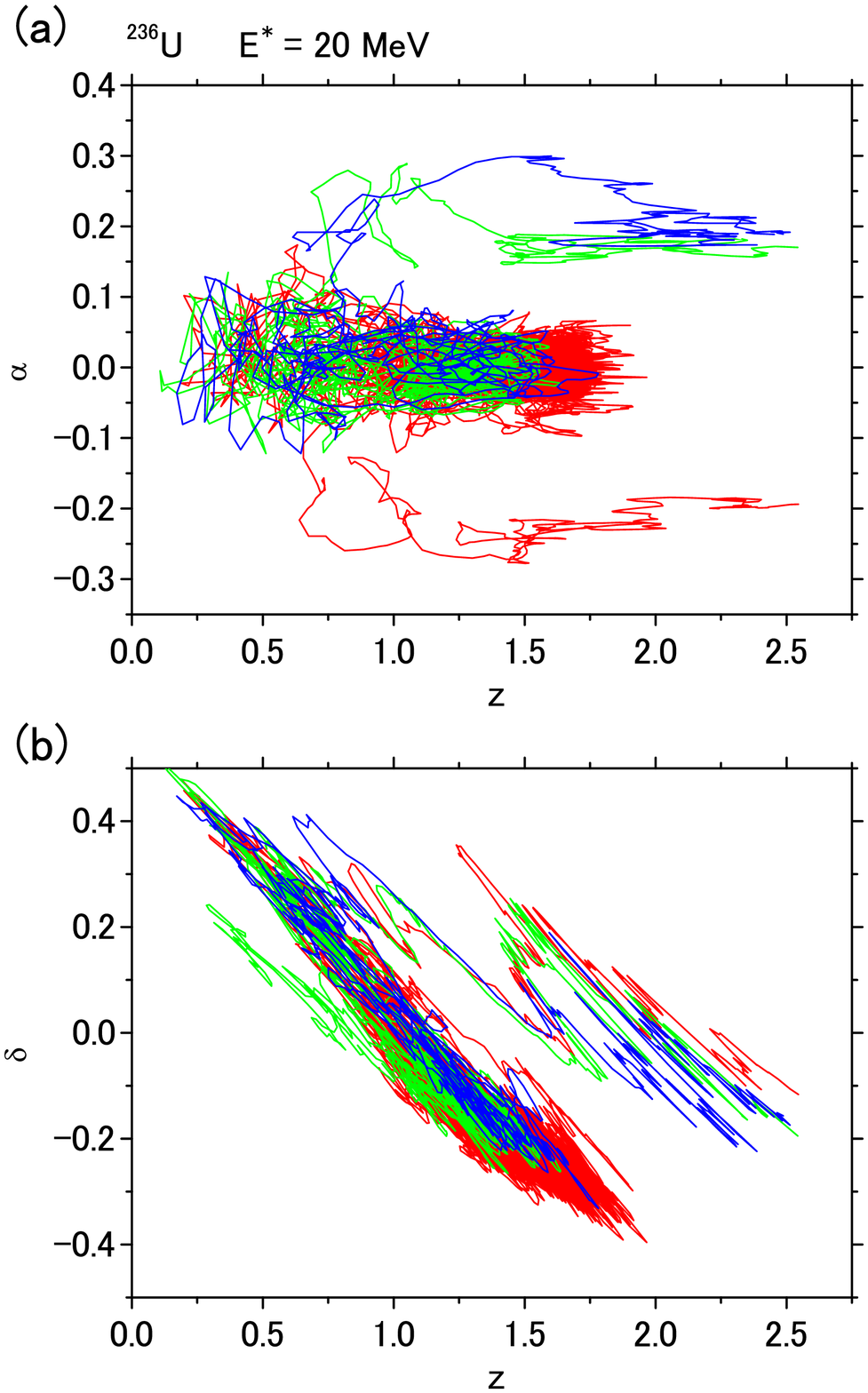}}
  \caption{Sample trajectories leading to the mass-asymmetric
fission region projected onto the $z$-$\alpha$
plane (a) and the $z$-$\delta$ plane (b), for the fission process of $^{236}$U at $E^{*}=20$ MeV.
The trajectories start at $\{z,\delta,\alpha\}=\{0.65,0.2,0.0\}$, where corresponds
to the second minimum of the potential energy surface.}
\label{fig_a1}
\end{figure}

Figure~\ref{fig_a1} shows sample trajectories to the mass-asymmetric
fission region and are projected onto the $z$-$\alpha$
plane (a) and $z$-$\delta$ plane (b) for a fission process of $^{236}$U at $E^{*}=20$ MeV.
Similarly to the calculation in reference \cite{arit13}, the trajectories start
at $\{z,\delta,\alpha\}=\{0.65,0.2,0.0\}$, which corresponds to the second minimum of
the potential energy surface.
The trajectories are initially trapped at the second pocket, then they escape from it.
Although they remain around this mass-symmetric region for a rather long time,
they even reach large $z$ values of  $z =1.5 - 1.75$.  However, they
do not move in a straight line to the separation region on the mass-symmetric fission path.
It can be observed that the trajectories leading to mass-asymmetric fission escape from the
region around $\{z,\alpha\} \sim \{0.8, \pm 0.2 \}$ in Fig.~\ref{fig_a1}(a).
On the $z$-$\delta$ plane in Fig.~\ref{fig_a1}(b), they also escape from
$\{z,\delta\} \sim \{0.8, 0.2\}$.
It appears that the trajectories leading to mass-asymmetric fission exhibit similar behavior
when they escape from the second pocket.

\begin{figure}
\centerline{
\includegraphics[height=.80\textheight]{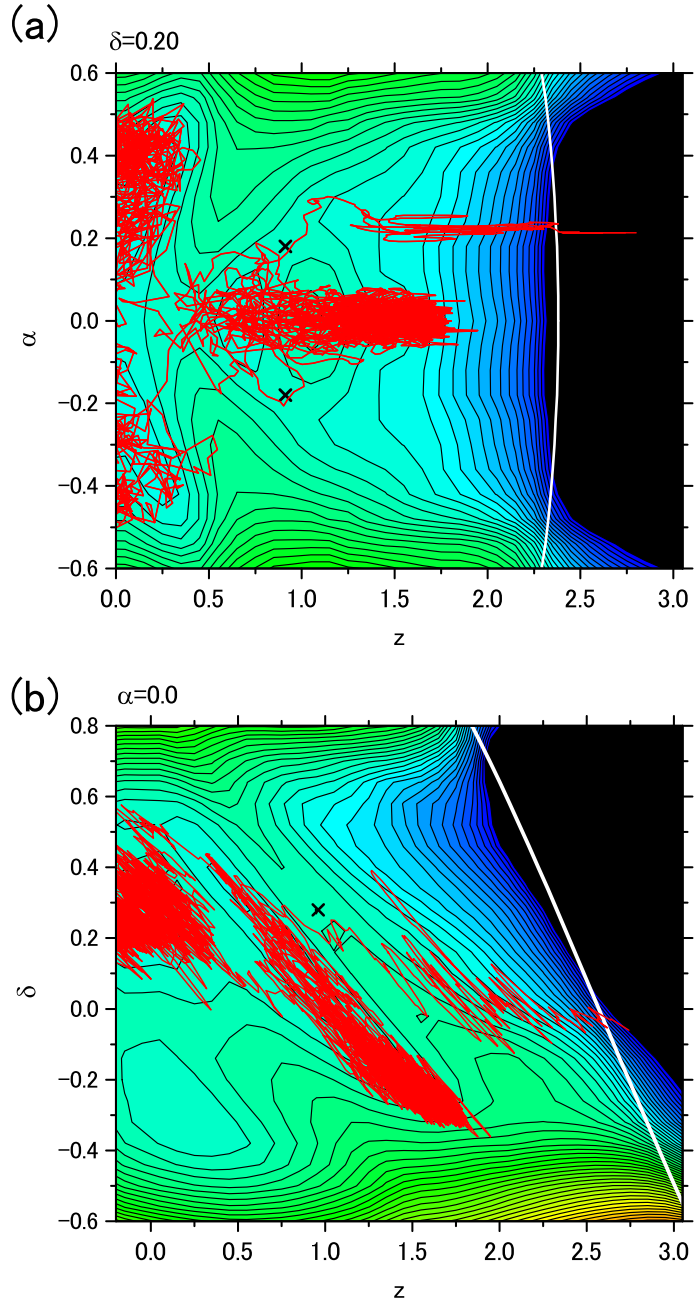}}
  \caption{Sample trajectory projected onto the $z$-$\alpha$
plane at $\delta = 0.2$ (a) and the $z$-$\delta$ plane $\alpha = 0.0$ (b) of
$V_{\rm LD}+E^{0}_{\rm shell}$ with $\epsilon = 0.35$ for $^{236}$U.
The trajectory starts
at the ground state $\{z, \delta, \alpha\} =\{0.0, 0.2, 0.0\}$ at $E^{*}=20$ MeV.
The fission saddle points are indicated by the symbol $\times$. The scission lines are
denoted by the white lines}
\label{fig_a2}
\end{figure}

\begin{figure}
\centerline{
\includegraphics[height=.40\textheight]{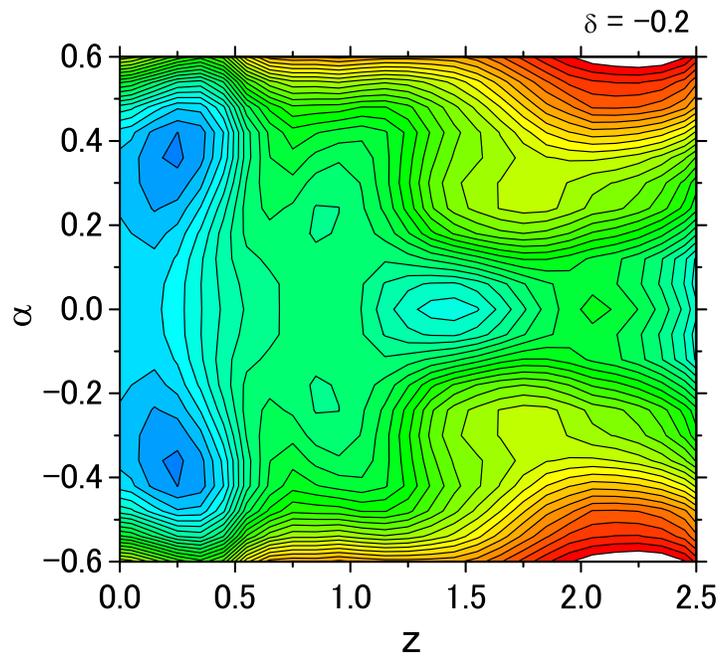}}
  \caption{The potential energy surface $V=V_{\rm LD}+E_{\rm shell}^{0}$ for $^{236}$U
in the $z$-$\alpha$ plane at $\delta =-0.2$ and $\epsilon = 0.35$.}
\label{fig_a3}
\end{figure}

To understand such behavior, we project the trajectories onto the potential energy surface.
Figure~\ref{fig_a2} shows a sample trajectory projected onto the $z$-$\alpha$
plane at $\delta = 0.2$ (a) and the $z$-$\delta$ plane at $\alpha = 0.0$ (b) of the potential
energy surface $V_{\rm LD}+E^{0}_{\rm shell}$ with $\epsilon = 0.35$ for $^{236}$U.
We define the scission point as the configuration with zero neck radius, as shown by a white line in Fig.~\ref{fig_a2}.
Here, as a demonstration, a trajectory starts
at the ground state $\{z, \delta, \alpha\} =\{0.0, 0.2, 0.0\}$ at $E^{*}=20$ MeV.
The trajectory remains at the first minimum and the second pocket for a long time.
The second pocket in Fig.~\ref{fig_a2}(b) corresponds to the pocket located around the
mass-symmetric region in the $z$-$\alpha$ plane shown in Fig.~\ref{fig_a3} with $\delta =-0.2$ and $\epsilon = 0.35$.
The trajectories in the mass-symmetric region in Figs.~\ref{fig_a1}(a) and \ref{fig_a2}(a) are trapped in the pocket
at the mass-symmetric region in Fig.~\ref{fig_a3}.
Then, they escape from the pocket and move along the valley that corresponds to $A \sim 140$.

In Fig.~\ref{fig_a2}, the fission saddle points are indicated by the symbol $\times$.
Even though the trajectories reach large $z$ values with $\alpha$=0, they do not move
to the scission region in a straight line.
Instead, the trajectories pass through the saddle points
before moving to the scission region.
We can state that the mass-asymmetric fission originates from the trajectories that overcome the fission saddle points,
which are located at the mass asymmetry corresponding to the position of the peak of the MDFF, where $A \sim 140$.

\begin{figure}
\centerline{
\includegraphics[height=0.80\textheight]{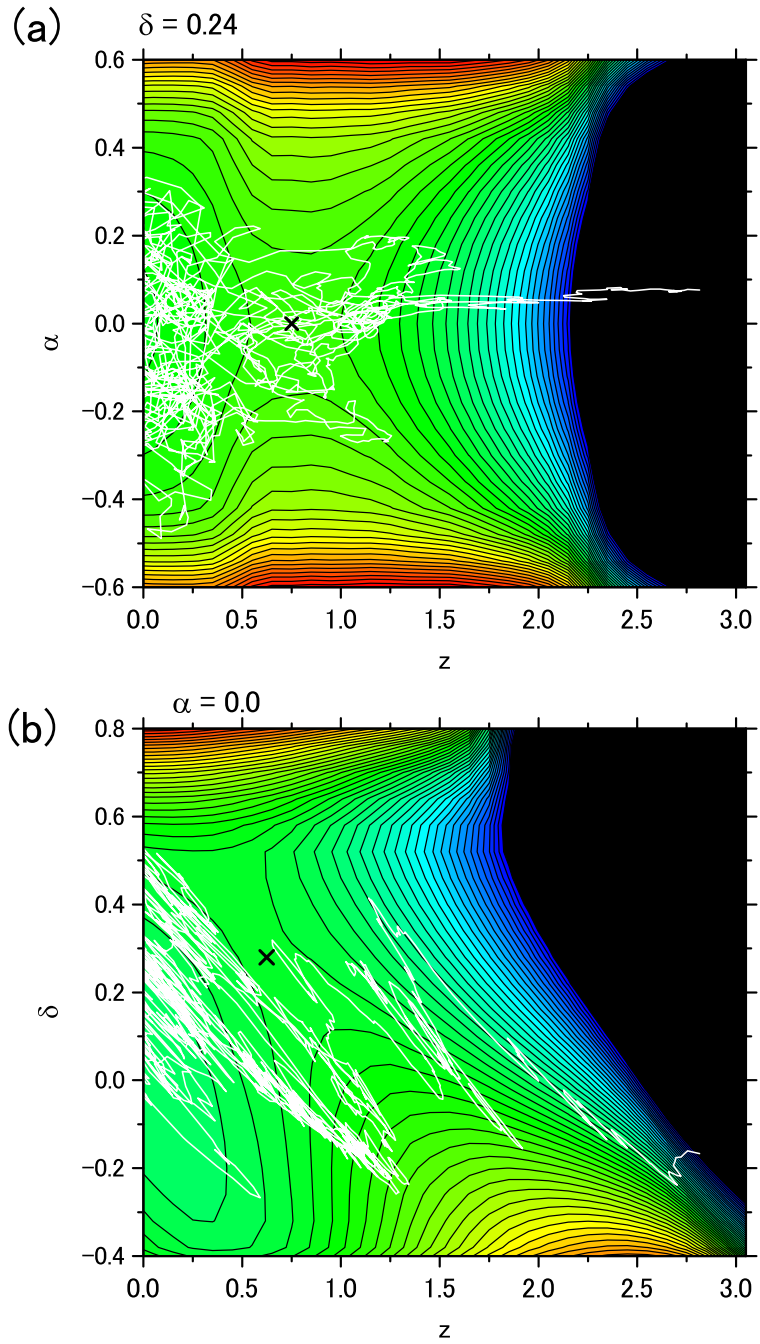}}
  \caption{Sample trajectory of the fission process of $^{236}$U at $E^{*}=20$ MeV
projected onto the $z$-$\alpha$ plane (a) and the $z$-$\delta$ plane (b) of $V_{\rm LD}$.
The fission saddle points are indicated by the symbols $\times$.}
\label{fig_a4}
\end{figure}

As a simple test, we calculate the MDFF using only the potential energy surface $V_{\rm LD}$, i. e., the lowest potential
energy is located in the mass-symmetric region.
Figure~\ref{fig_a4} shows a sample trajectory projected onto the $z$-$\alpha$ plane with $\delta=0.24$ (a) and the
$z$-$\delta$ plane with  $\alpha = 0.0$ (b) of the potential energy surface $V_{\rm LD}$.
On the potential energy surface in Fig.~\ref{fig_a4}, the fission saddle points are indicated by the symbol $\times$.
The trajectory overcomes such saddle points and moves to the mass-symmetric fission region.
Such behavior is very different from that on  $V_{\rm LD}+E^{0}_{\rm shell}$.
We conclude that the destiny of the trajectory, i. e., whether it moves
to the mass-symmetric fission region or mass-asymmetric region, is decided at the fission saddle point.

In Fig.~\ref{fig_a2}, the fission saddle point is located in the mass-asymmetric region.
Thus, we investigate the potential landscape on the $z$-$\delta$ plane at $\alpha=0.0$, 0.18, and 0.24, as shown in Figs.~\ref{fig_a5}.
Compared with the case of $\alpha=0.0$, the fission barrier height is lower for $\alpha=0.18$ and it disappears for $\alpha=0.24$.
In these cases, the trajectory easily overcomes the fission saddle point in the $z$-$\delta$ plane, and finally it
moves to the mass-asymmetric fission region.
As a result of the shell correction energy, the barrier height at the mass-asymmetric fission saddle point with $\alpha \sim 0.2$
becomes lower than that in the other mass-asymmetric fission region.
This is why the trajectories move to the mass-asymmetric fission region.


\section{Width of peak in MDFF and fission dynamics} 

Upon investigating the behavior of the trajectories more precisely,
we notice that the trajectory on the $z$-$\delta$ plane fluctuates in the direction of -45$^\circ$, as shown in Figs.~\ref{fig_a1}
and \ref{fig_a2}.
We project other sample trajectories of the fission process of $^{236}$U at $E^{*}=20$ MeV
onto the $z$-$\alpha$ plane with $\delta=0.2$ (a) and onto the $z$-$\delta$ plane
with $\alpha=0.18$ (b) in Fig.~\ref{fig_a6}; these trajectories start at the second pocket and escape from it in a rather short time.
Even after they overcome the fission saddle point, such oscillations are observed up to the scission point.
In general, the trajectories move along the slope of a potential energy surface.
To understand the trajectory behavior clearly, we discuss it on the potential energy surface $V_{\rm LD}$.
Figure \ref{fig_a7} shows the trajectory without the random force on $V_{\rm LD}$, projected onto the $z$-$\alpha$ plane
with $\delta=0.24$ (a) and onto the $z$-$\delta$ plane with $\alpha=0.0$ (b).
The trajectories start at the saddle point and move down to the separation region along the potential slope owing to the drift force $-\frac{\partial V}{\partial q_{i}}$ in Eq.~(\ref{lange1}).
However, when we take into account the random force, the trajectory on the $z$-$\delta$ plane shows an oscillation in the direction of -45$^\circ$.
We have investigated this special directivity and revealed the reason for it.
It originates from the characteristic of  the friction tensor, mainly the nondiagonal terms, via the Einstein relation.
This will be explained in detail in a forthcoming paper.


In addition to the origin of the directivity, the oscillation of the trajectory is very important in fission dynamics.
After the trajectories overcome the fission saddle point, as shown in Figs.~\ref{fig_a1}(b),
 \ref{fig_a2}(b) and \ref{fig_a6}(b), they fluctuate frequently and move down the potential slope step by step because
the direction of the oscillation is not perpendicular to the contour of the potential energy surface.
In the region around $z > 1.5$, $\delta < 0$, roughly speaking, the contour of the potential is parallel
to the $z-$axis; therefore, the
trajectory climbs and descends the potential slope as a result of the random force and drift force, respectively.
This corresponds to the thermal fluctuation of the nuclear shape
around the scission point, as described in Fig.~\ref{fig_a8}.
The nuclear shape with $\delta < 0$ corresponds to the oblate shape, but actually it indicates that
the curvature of the edge side is negative, which is shown by the gray line
in Fig.~\ref{fig_a8} corresponding to the nuclear shape at $\{z,\delta,\alpha\}=\{2.5,-0.2,0.2\}$.
This means that nuclear fission does not occur with continuous stretching, as exhibited by starch syrup.
Around the scission point, during the shape vibration of the length and breadth of the fissioning fragments,
the nucleus is split suddenly by a strong vibration of the length ($-\delta$ direction),
which reduces the density in the neck region \cite{iwam13}.
We can conclude that the widths of the MDFF around the peaks are determined by such fluctuations near the scission point.
Since the calculated MDFF shows good agreement with the experimental data in reference \cite{arit13},
it supports the hypothesis that the vibration of the nuclear shape is essential to describe nuclear fission correctly.

\begin{figure}
\centerline{
\includegraphics[height=.80\textheight]{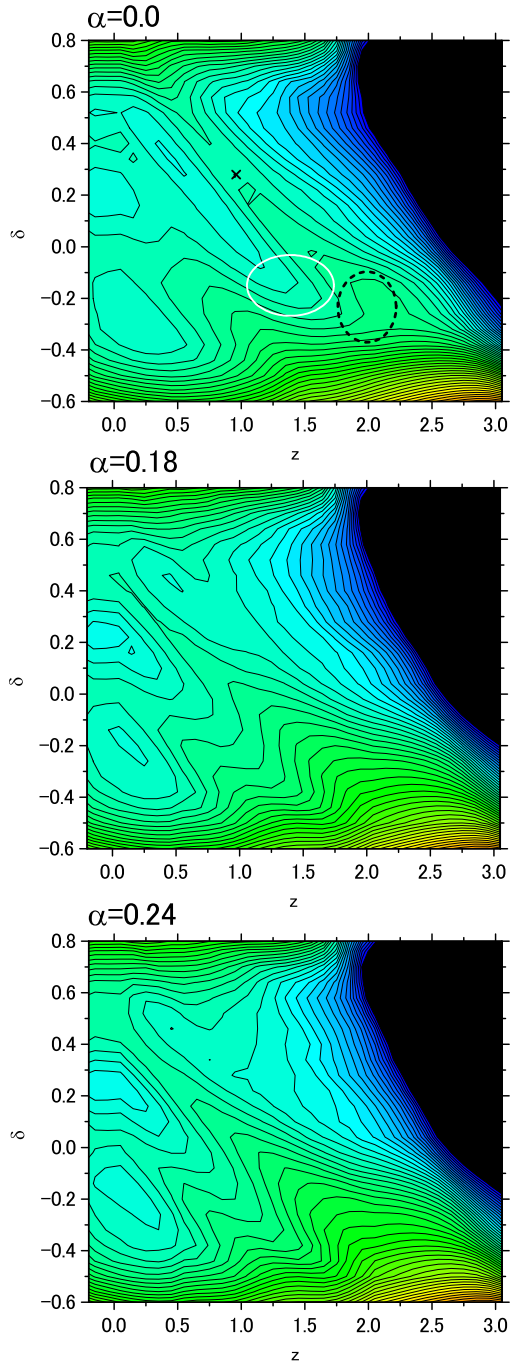}}
  \caption{The potential energy surface $V=V_{\rm LD}+E_{\rm shell}^{0}$ for $^{236}$U with $\epsilon = 0.35$
in the $z$-$\delta$ plane at $\alpha =0.0$, 0.18, and 0.24.
The white circle indicates the second pocket where the trajectories remain deep in the negative $\delta$ region.
Symmetric fission barrier is denoted by a dashed circle.}
\label{fig_a5}
\end{figure}
\begin{figure}
\centerline{
\includegraphics[height=.80\textheight]{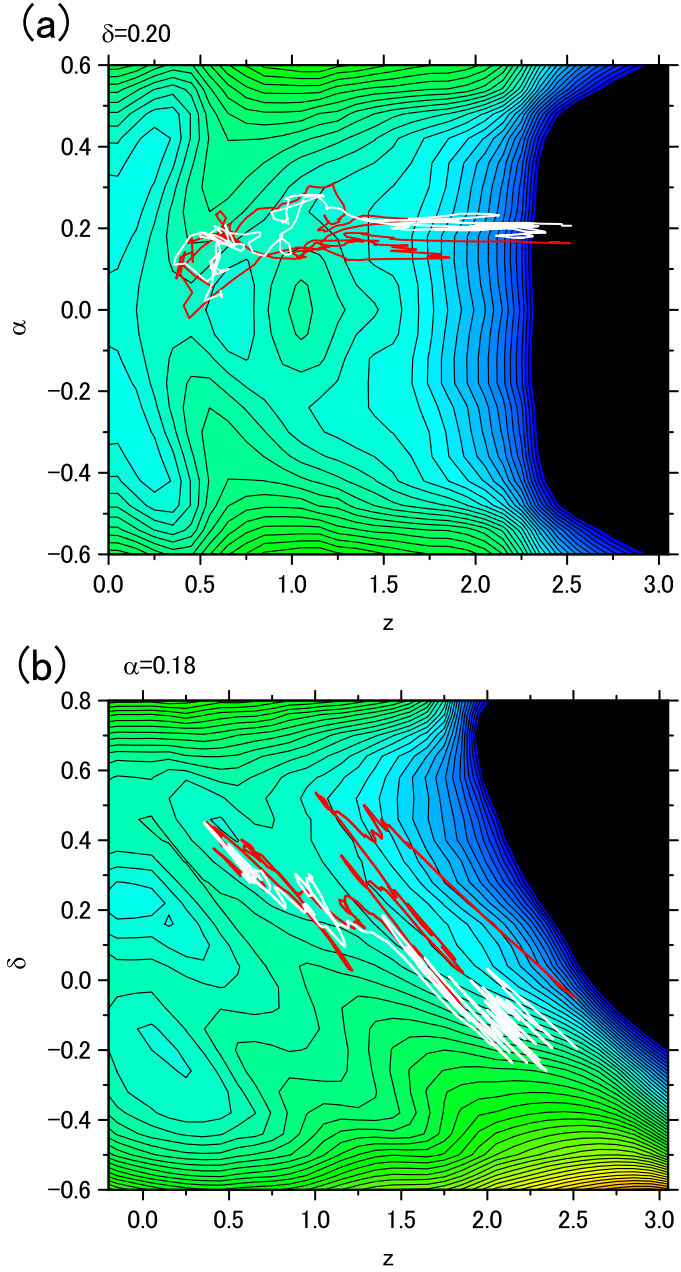}}
  \caption{Sample trajectories of the fission process of $^{236}$U at $E^{*}=20$ MeV are projected
onto the $z$-$\alpha$ plane with $\delta=0.2$ (a) and the $z$-$\delta$ plane with $\alpha=0.18$ (b).
The trajectories start at the second pocket and escape from it in a rather short time.}
\label{fig_a6}
\end{figure}
\begin{figure}
\centerline{
\includegraphics[height=.80\textheight]{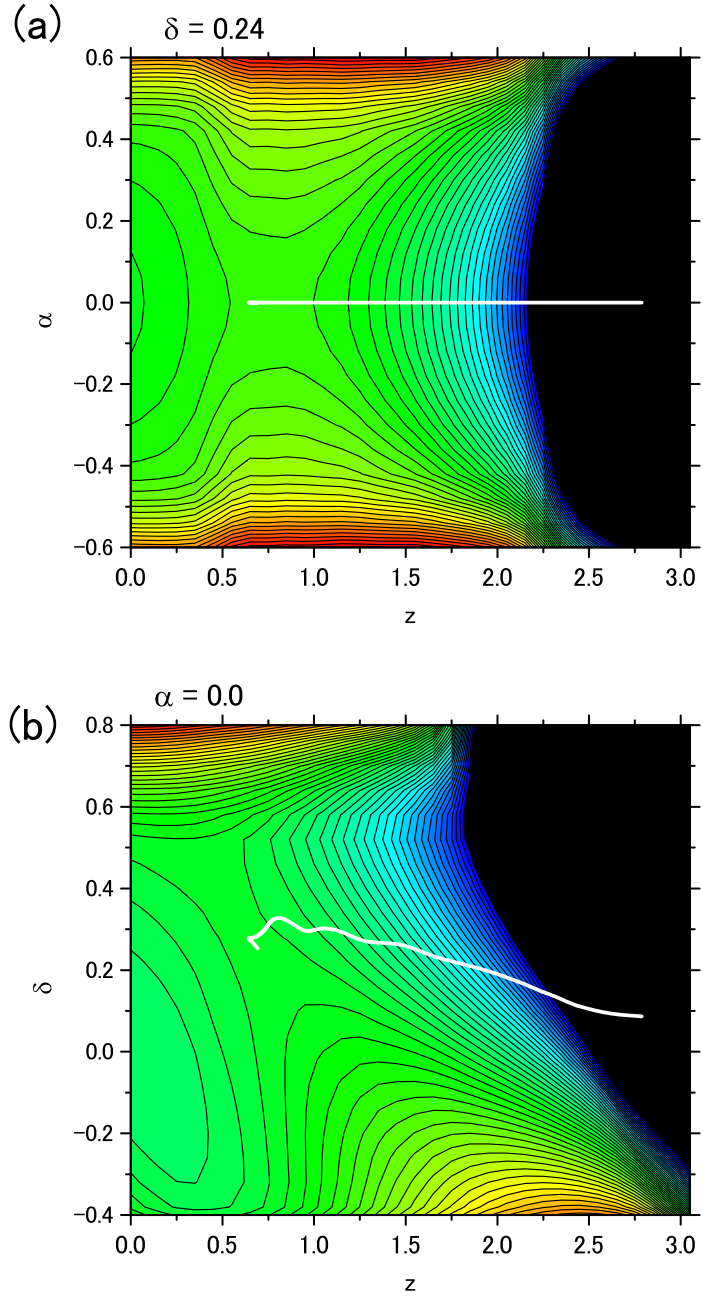}}
  \caption{The trajectory of the fission process of $^{236}$U at $E^{*}=20$ MeV,  without
the random force on $V_{\rm LD}$, projected onto the $z$-$\alpha$ plane
with $\delta=0.24$ (a) and  the $z$-$\delta$ plane with $\alpha=0.0$ (b).
The trajectories start at the saddle point.}
\label{fig_a7}
\end{figure}
\begin{figure}
\centerline{
\includegraphics[height=.40\textheight]{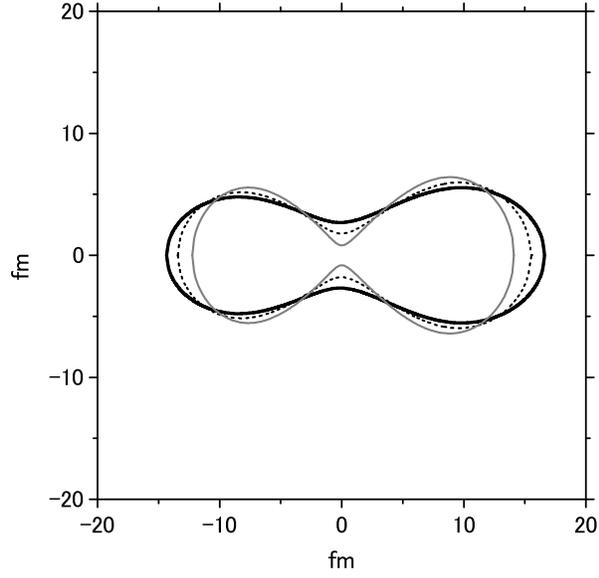}}
  \caption{Nuclear shapes around the scission point of $^{236}$U at $E^{*}=20$ MeV.
The gray line corresponds to the nuclear shape at $\{z,\delta,\alpha\}=\{2.5,-0.2,0.2\}$.}
\label{fig_a8}
\end{figure}

In addition, in Figs.~\ref{fig_a1} and ~\ref{fig_a2}, the trajectories are trapped
in the second pocket and remain deep in the negative $\delta$ region.
This region is indicated by the white circle in Fig.~\ref{fig_a5} with $\alpha=0$.
The shape evolution with a negative $\delta$ is very important in fission dynamics, as discussed above.
In Fig.~\ref{fig_a1}(a), although the trajectories are trapped in the pocket around $\alpha = 0$ and
even reach the region with large $z$-values ($z \sim 1.75$), they do not escape along the mass-symmetric path.
In Fig. \ref{fig_a5}(a), a barrier can be observed between $z=1.75$ and 2.0 with $\delta=-0.2$, which we call the ``symmetric fission barrier" and indicate by a dashed circle.
Owing to the symmetric fission barrier, the trajectory does not escape along the
mass-symmetric fission valley in Fig.~\ref{fig_a3}.

It is well known that the MDFF of Fm and Th isotopes changes abruptly between mass-asymmetric and symmetric fission modes depending on the number of neutrons.
Even a difference of one neutron in some isotopes causes a marked change in the MDFF.
The mechanism of this phenomenon is not yet understood clearly.
From our trajectory analysis, we may conclude that whether mass-symmetric or asymmetric fission occurs is very sensitive
to the height of this mass-symmetric fission barrier.
When the mass-symmetric barrier height is lower and the trajectories can overcome it,
mass-symmetric fission becomes dominant and it may produce a very sharp single peak.
This is realized when the potential energy surface is described correctly by the appropriate
nuclear shape variables; obtaining such a description is planned as a future work.


The MDFF is affected by the behavior of the trajectory on the potential energy surface.
The width of the peaks of the MDFF is determined by the oscillation of trajectories
after overcoming the fission saddle point.
The behaviors of trajectories can be categorized roughly into two regions in the $z$-$\delta$ plane: the positive $\delta$
and the negative $\delta$ regions.
In the positive $\delta$ region as shown in Fig.~\ref{fig_a6}(b), the contour lines of the potential energy surface
are in the -45$^\circ$ direction, which coincides with the direction of the oscillation of the trajectory indicated in red (gray in monotone print).
Therefore, the trajectory quickly moves down along the contour line from a large positive $\delta$
to a small $\delta$, in the $z$-$\delta$ plane. This means that these is no fluctuation in the $\alpha$ direction
in the $z > 1.75$ region on the $z$-$\alpha$ plane in Fig.~\ref{fig_a6}(a).
On the other hand, in the negative $\delta$ region, because of the barrier,
the trajectory shows oscillates in the -45$^\circ$ direction in the $z$-$\delta$ plane, as shown by the white trajectory.
This corresponds to the fluctuation of the $\alpha$ parameter around the scission line in the $z$-$\alpha$ plane.
The widths of the peaks of the MDFF are determined by the potential landscape around the scission line,
which is mainly in the negative $\delta$ region on the $z$-$\delta$ plane.



\section{Roles of friction tensor in fission process}

In our previous study \cite{arit13}, we investigated how the MDFF is affected by the strength of the friction tensor.
The calculated results depended slightly on  the strength of the friction tensor.
However, in the fusion-fission process in the superheavy mass region, such dependence is rather strong \cite{arit06,arit11}.
Moreover, even in the fission process at a high excitation energy, the MDFF and TKE are
affected by the value of the friction tensor \cite{karp01}.
Here, we investigate the reason for this conflict.


The friction in the Langevin equation strongly affects the dissipation of the kinetic energy into the intrinsic energy.
In the fusion process, this effect is very prominent \cite{arit09,arit11}.
On the other hand, to treat the fission process at a low excitation energy, we assume that thermal equilibrium has
already been reached.
Therefore, we perform the calculation at the starting point where all the kinetic energy has already been transferred
into the intrinsic energy.

\begin{figure}
\centerline{
\includegraphics[height=.50\textheight]{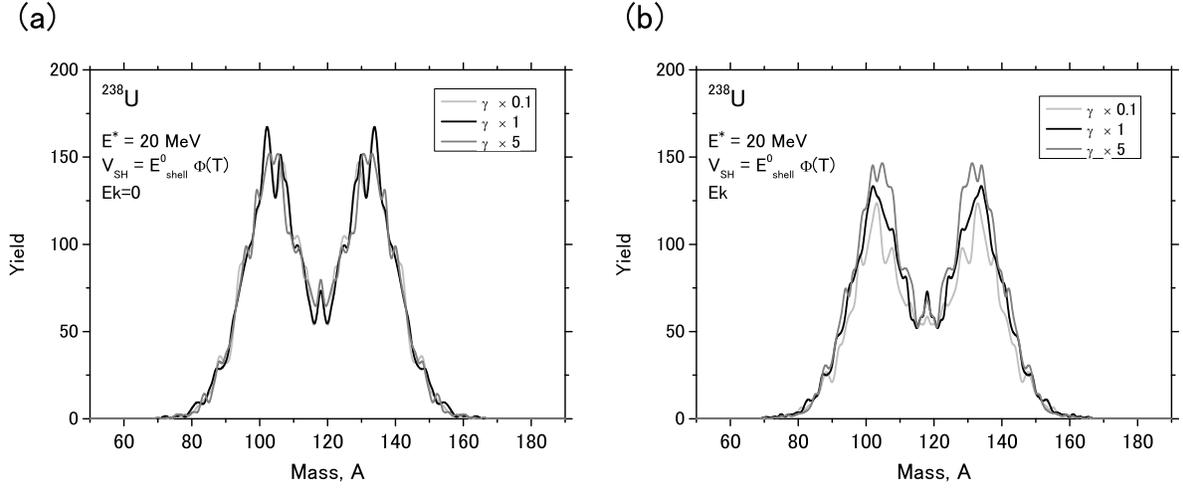}}
  \caption{MDFF of $^{236}$U at $E^{*}=20$ MeV for the friction tensor $\gamma$ multiplied by 0.1, 1, and 5.
The temperature dependence of shell correction energy is considered as $E_{d}=20$ MeV.
(a) and (b) are obtained without and with the initial boost in the $-z$ direction, respectively.}
\label{fig_a9}
\end{figure}

To determine the effect of the friction tensor in the fission process,
we give the system a momentum in the $-z$ direction at the starting point located at
the second minimum on the potential energy surface.
Figure~\ref{fig_a9} shows the MDFF of $^{236}$U at $E^{*}=20$ MeV for the friction tensor multiplied by 0.1, 1, and 5.
Figures~\ref{fig_a9}(a) and (b) were obtained without and with the initial boost
in the $-z$ direction, respectively.
Although we could not observe any dependence of the MDFF on the friction tensor in Fig.~\ref{fig_a9}(a),
such a dependence appears in Fig.~\ref{fig_a9}(b).
Therefore, the rate of dissipation of the kinetic energy affects the MDFF.
In a low-energy case, if the kinetic energy of the system has already dissipated or thermal equilibrium
has been reached, the friction dependence of the MDFF does not appear.

\begin{figure}
\centerline{
\includegraphics[height=.50\textheight]{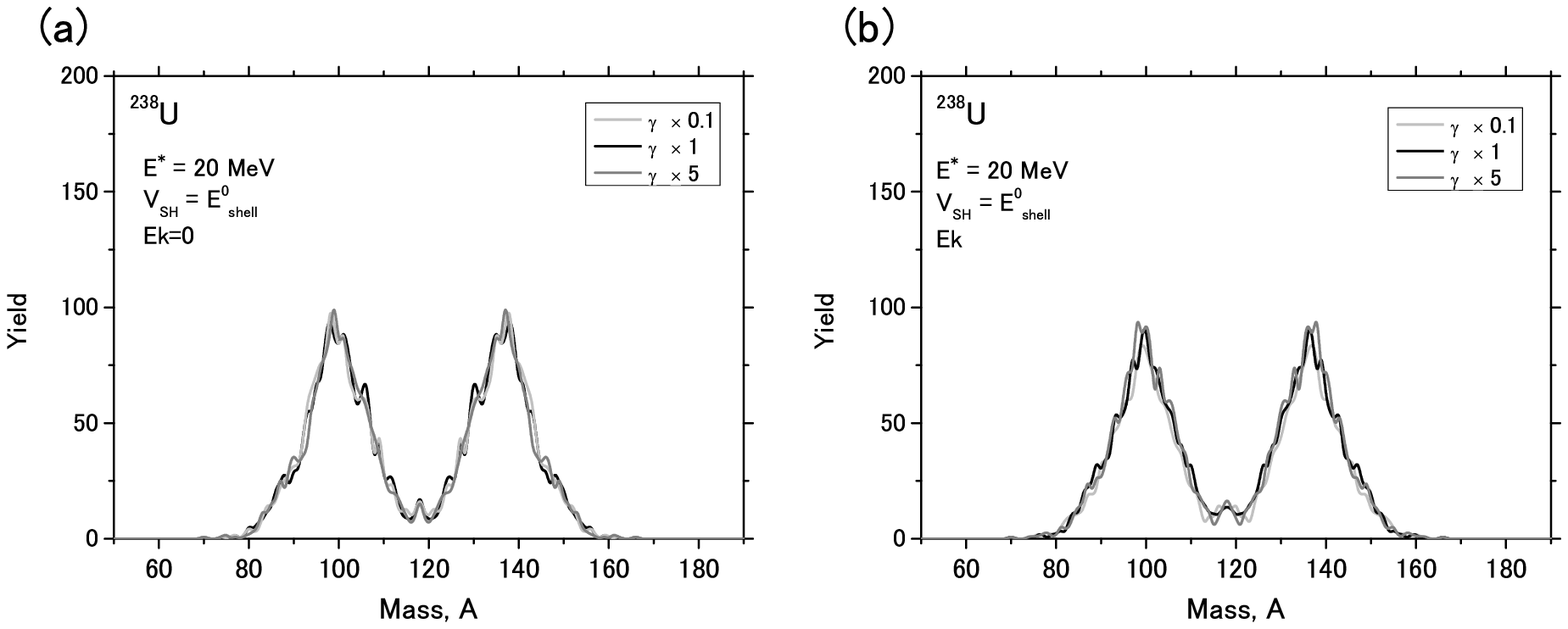}}
  \caption{MDFF of $^{236}$U at $E^{*}=20$ MeV for the friction tensor $\gamma$ multiplied by 0.1, 1, and 5.
The temperature dependent of shell correction energy is not considered, namely, $V_{\rm LD}+E^{0}_{\rm shell}$.
(a) and (b) are obtained without and with the initial boost in the $-z$ direction, respectively.}
\label{fig_a10}
\end{figure}

Moreover, we carry out the same calculation but with the temperature dependence of the shell correction energy removed.
We use the potential energy surface $V_{\rm LD}+E^{0}_{\rm shell}$.
The results for the MDFF without and with the initial momentum in the $-z$ direction are shown in Figs.~\ref{fig_a10}(a) and (b),
respectively.
We could not find any dependence on the friction tensor in either case.
Because of the strong shell structure, the trajectories are trapped deep in the pockets at the ground state or the
second minimum.
While the trajectories remain in the pocket, their kinetic energy dissipates inside the fission barrier, meaning that it does
not affect the MDFF.

\begin{figure}
\centerline{
\includegraphics[height=.40\textheight]{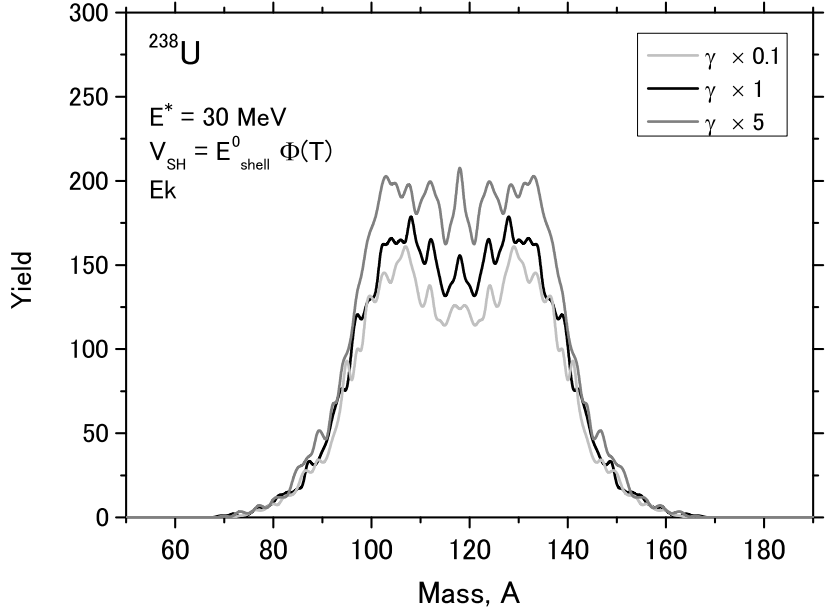}}
  \caption{MDFF of $^{236}$U at $E^{*}=30$ MeV for the friction tensor $\gamma$ multiplied by 0.1, 1, and 5.
The temperature dependence of shell correction energy is considered as $E_{d}=20$ MeV.
At the staring point, the system has the initial boost in the $-z$ direction.}
\label{fig_a11}
\end{figure}

The effects of the friction tensor will be stronger when the system has a higher kinetic energy.
Under the conditions used in Fig.~\ref{fig_a9}(b) but with $E^{*}=30$ MeV, we calculate the MDFF,
as shown in Fig.~\ref{fig_a11}.
The dependence on the friction tensor is more prominent than in the case of $E^{*}=20$ MeV.

\section{Nuclear shape at scission point and Total Kinetic Energy}

\begin{figure}
\centerline{
\includegraphics[height=.40\textheight]{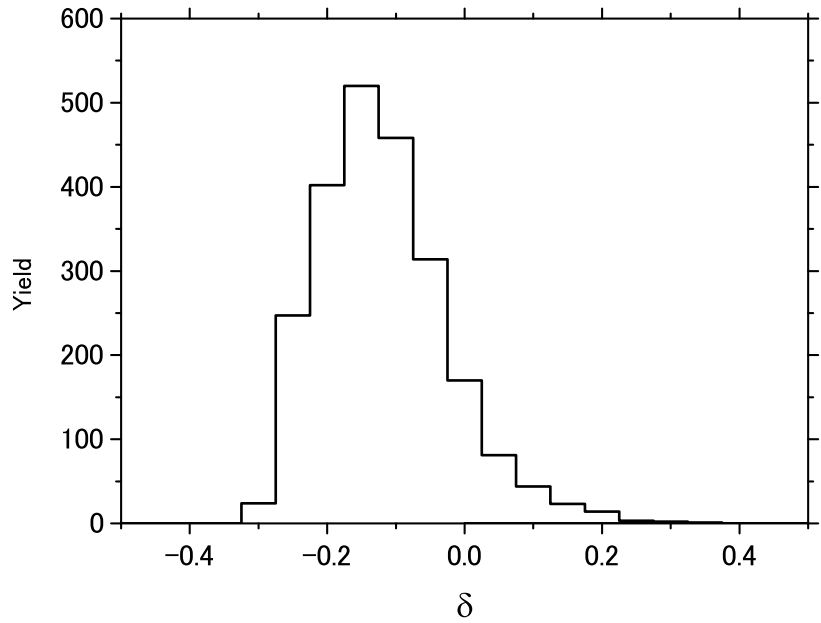}}
  \caption{The distribution of the deformation parameter $\delta$ at the scission point
for the fission of $^{236}$U at $E^{*}=20$ MeV.}
\label{fig_a12}
\end{figure}

In Sec.~IV, we pointed out that a nuclear shape with a negative $\delta$ is very important in fission dynamics,
particularly around the scission point.
The movement of trajectories in the negative $\delta$ direction driven by the random force leads to the splitting
into fragments.
The nuclear shapes around the scission point are presented in Fig.~\ref{fig_a8}.
The shape denoted by the gray line is similar to that obtained using the statistical scission model in Fig.~5 of
reference \cite{fong56}.
The distribution of the deformation parameter $\delta$ at the scission point
for the fission of $^{236}$U at $E^{*}=20$ MeV is shown in Fig.~\ref{fig_a12}.
Fragments with negative values of $\delta$ are dominant.

To clarify the configuration at the scission point, we investigate the TKE of the fission fragments.
We calculated the average TKE of the fission fragments $\langle TKE \rangle$ of $^{236}$U at $E^{*}=20$ MeV.
We obtained $\langle TKE \rangle = 171.8$ MeV, which is in agreement with the experimental data \cite{vand73}.
Also, the dependence of $\langle TKE \rangle$ on the mass number of the fission fragments was calculated, which was shown in Fig.~7
of reference \cite{arit13}.
Because of this agreement with the experimental data for the TKE, we conclude that the configuration
at the scission point is compact, such as that shown by the gray line in Fig.~\ref{fig_a8}.

\begin{figure}
\centerline{
\includegraphics[height=.40\textheight]{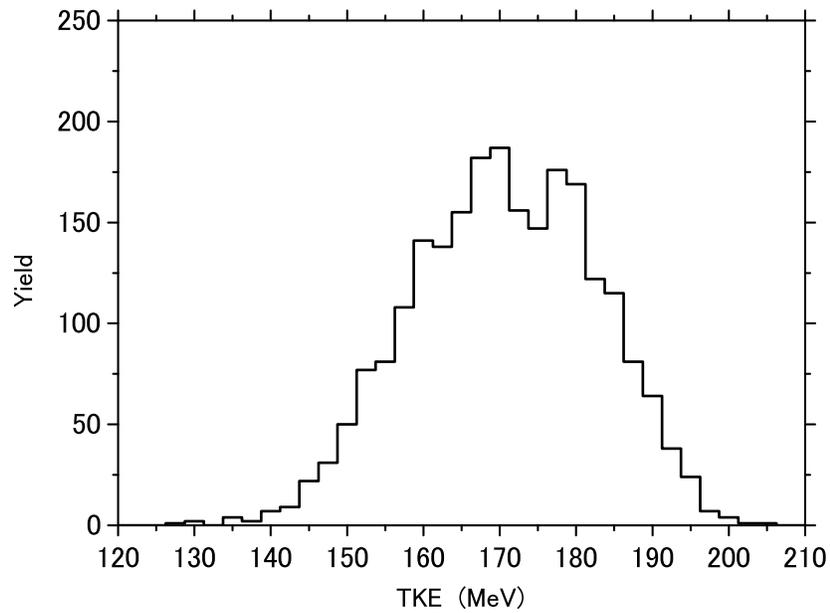}}
  \caption{The TKE distribution of the fission fragments of $^{236}$U at $E^{*}=20$ MeV.}
\label{fig_a13}
\end{figure}

\begin{figure}
\centerline{
\includegraphics[height=.40\textheight]{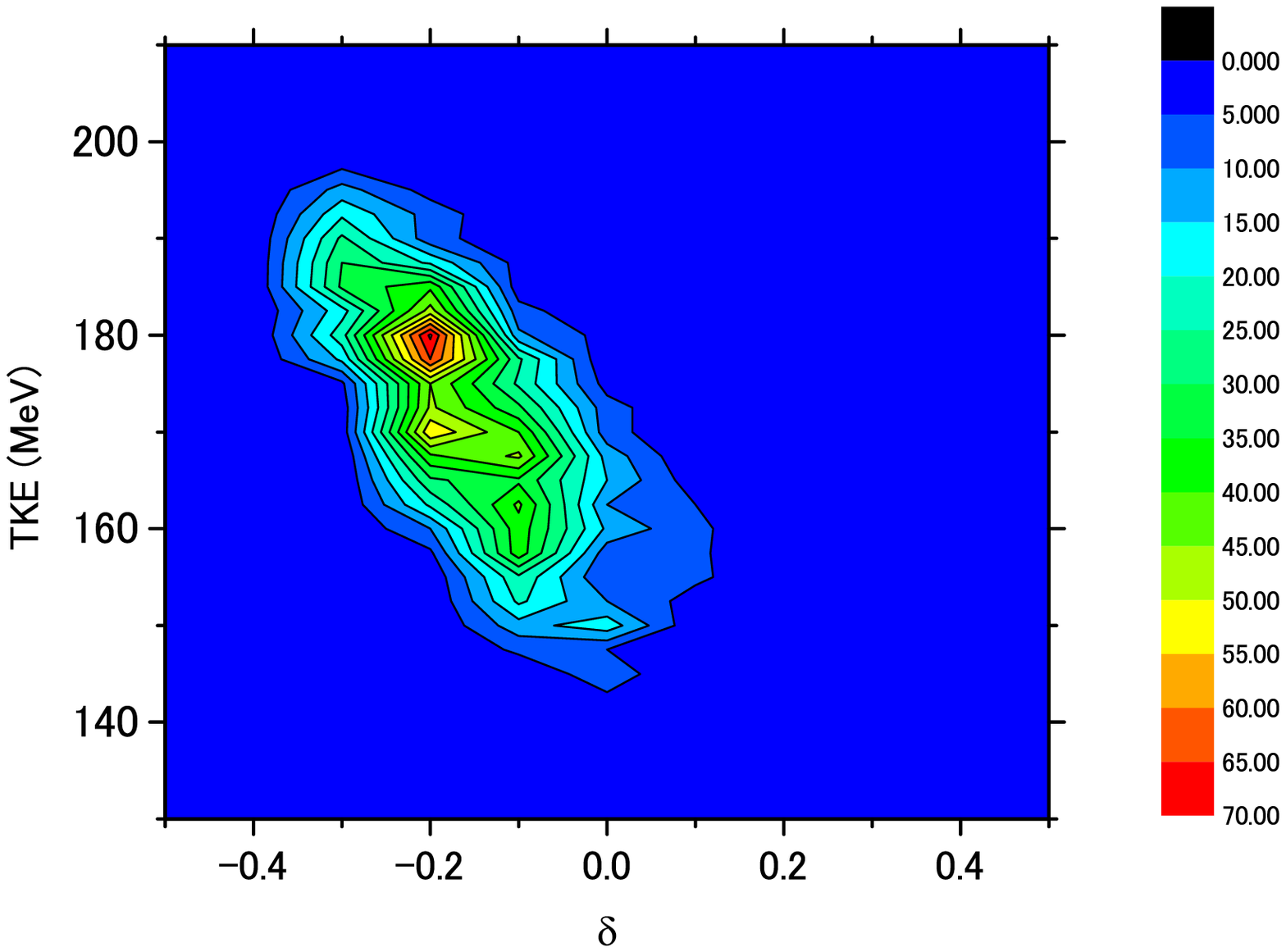}}
  \caption{The correlation between the TKE and the parameter $\delta$ of $^{236}$U at $E^{*}=20$ MeV.}
\label{fig_a14}
\end{figure}

The TKE distribution of the fission fragments of this system is shown in Fig.~\ref{fig_a13}.
The distribution is approximately Gaussian.
Figure \ref{fig_a14} shows the correlation between the TKE and the parameter $\delta$.
Since the configuration with a negative $\delta$ corresponds to the compact shape, the TKE of such fragments is higher than that
of fragments with a positive $\delta$.
The fissioning fragments with the compact configuration are dominant in this system.

\section{Summary}

In this paper, we investigated the fission process at a low excitation energy using the Langevin equation.
By analyzing the trajectories in our model \cite{arit13},
we clarified the origin of the mass-asymmetric fission events of uranium at a low excitation energy.
This is the first time a clear explanation for the mass-asymmetric fission has been given by a dynamical approach,
even though mass-asymmetric fission was discovered more than 70 years ago.

The MDFF of $^{236}$U at $E^{*}=20$ MeV showed mass-asymmetric fission.
We found that the origin of the peak positions is mainly related to the positions of the fission saddle points,
which are caused by the shell correction energy.
To escape from the potential pockets around the ground state or the second minimum, almost all the
trajectories pass through the fission saddle points and move to the mass-asymmetric fission region.
After overcoming the fission saddle points, the trajectories fluctuate frequently owing to the random force
in the Langevin equation and approach the scission point.
The fluctuation around the scission point determines the widths of the peaks of the MDFF.

By analyzing the fission process and investigating the shape evolution precisely, we found
that the movement in the negative $\delta$ direction around the scission point is essential for the fission process.
We stress that nuclear fission does not occur with continuous stretching, such as that observed in starch syrup.
Rather, it occurs around the scission point; during the shape vibration of the length and breadth of the fissioning fragments,
the nucleus is suddenly split by a strong vibration in the negative $\delta$ direction.
Such a mechanism in fission dynamics and the configuration with negative $\delta$ values at the scission point are supported by the fact that the calculated MDFF and TKE show  good agreement with the experimental data
in reference \cite{arit13}.
In addition, we pointed out that the trajectories do not always move along the minimum points on the potential energy surface owing to the random force, nor fluctuate around the trajectory without the random force (mean trajectory).
Although analyses of the fission process using the static potential energy surface are sensible in some cases, it is not enough to revel the complicated fission process.
In this paper, we thus presented new viewpoints of the mechanism of fission dynamics.


As further study, we plan to improve the model by increasing the number of variables,
at least by introducing independent deformation parameters
for each fragment. 
Moreover, the neutron emission from the fissioning system and also from fission fragments should be included in the model.
Though such improvements of the model, we aim to decrease the differences between the calculated
MDFF and the experimental data.



\section*{Acknowledgments}

Present study is the results of ``Comprehensive study of delayed-neutron yields for accurate
evaluation of kinetics of high-burn up reactors"
entrusted to Tokyo Institute of Technology by the Ministry of Education, Culture, Sports, Science and Technology of Japan (MEXT).
The authors are grateful to Dr. A.~Iwamoto, Prof. M.~Ohta, Prof. T.~Wada, Dr. K.~Nishio,
Dr.~A.V.~Karpov, Dr.~F.A.~Ivanyuk and Prof.~V.I.~Zagrebaev
for their helpful suggestions and valuable discussions.
Special thanks are deserved to Mr.~K.~Hanabusa (HPC Systems Inc.)  for his technical supports
to operate the high performance computer.


\newpage

\vspace{10cm}
\newpage
\clearpage

\vspace{10cm}
\newpage
\clearpage


\begin{thebibliography}{0}
\bibitem{harn39} O.~Hahn and F.~Stra{\ss}mann, Naturwissenschaften {\bf 27}, 11 (1939).
\bibitem{meit39} L. Meitner and O.R.~Frisch, Nature (London) {\bf 143}, 239 (1939).
\bibitem{bohr39} N.~Bohr and J.A.~Wheeler, Phys. Rev. {\bf 56}, 426 (1939).
%
\bibitem{arit13} Y.~Aritomo and S.~Chiba, arXiv:1305.6132, accepted in Phys. Rev. C.
%
\bibitem{arit04} Y.~Aritomo and M.~Ohta, Nucl. Phys. A {\bf 744}, 3 (2004).
\bibitem{maru72}J.~Maruhn and W.~Greiner, Z. Phys. {\bf 251}, 431 (1972).
\bibitem{sato78}K.~Sato, A.~Iwamoto, K.~Harada, S.~Yamaji, and S.~Yoshida,
 Z. Phys. A {\bf 288}, 383 (1978).
\bibitem{krap79}H.J.~Krappe, J.R.~Nix, and A.J.~Sierk, Phys. Rev.
C {\bf 20}, 992 (1979).
\bibitem{igna75} A.V.~Ignatyuk, G.N.~Smirenkin, and A.S.~Tishin, Sov.
J. Nucl. Phys. {\bf 21}, 255 (1975).
\bibitem{bloc78}J.~Blocki, Y.~Boneh, J.R.~Nix, J.~Randrup, M.~Robel,
A.J.~Sierk, and W.J.~Swiatecki, Ann. Phys. {\bf 113}, 330 (1978).
\bibitem{nix84}J.R.~Nix and A.J.~Sierk, Nucl. Phys. A {\bf 428}, 161c (1984).
\bibitem{feld87}H.~Feldmeier, Rep. Prog. Phys. {\bf 50}, 915 (1987).
\bibitem{wada93} T.~Wada, Y.~Abe and N.~Carjen, Phys.~Rev.~Lett.~{\bf 7}, 3538(1993).
%
\bibitem{davi76}K.T.R.~Davies,A.J.~Sierk, and J.R.~Nix, Phys. Rev. C {\bf 13}, 2385 (1976).

\bibitem{iwam13} A.~Iwamoto, and F.A.~Ivanyuk, private communication.
\bibitem{arit06} Y.~Aritomo, Nucl. Phys. A {\bf 780}, 222 (2006).
\bibitem{arit11} Y.~Aritomo, S.~Chiba and K.~Nishio, Phys. Rev. C {\bf 84}, 024602 (2011).
\bibitem{karp01} A.V.~Karpov, P.N.~Nadtochy, D.V.~Vanin, and G.D.~Adeev, Phys. Rev. C {\bf 63}, 054610 (2001).
\bibitem{arit09} Y.~Aritomo, Phys. Rev. C {\bf 80}, 064604 (2009).
\bibitem{fong56} P.~Fong, Phys. Rev. {\bf 102}, 434 (1956).
\bibitem{vand73} R.~Vandenbosh and J.R.~Huizenga, {\it Nuclear Fission}, Academic Press, New York and London (1973).

\end{thebibliography}
\end{document}